\documentclass[12pt]{article}
\usepackage{amsmath}
\usepackage{graphicx}
\usepackage{hyperref}
\usepackage{lmodern}
\usepackage{amssymb}
\usepackage{booktabs}

\usepackage{feynmf}

 \newcommand{\be}{\begin{equation}}
\newcommand{\ee}{\end{equation}}

\newcommand{\beqa}{\begin{eqnarray}}
\newcommand{\eeqa}{\end{eqnarray}}
\newcommand{\nn}{\nonumber}

\newcommand{\<}{\langle}

\def\CA {{\cal A}}

\def\CD {{\cal D}}

\def\CG {{\cal G}}

\def\CL {{\cal L}}

\def\CP {{\cal P}}

\def\CV {{\cal V}}

\begin{document}
\setlength{\unitlength}{1mm}
\setlength{\baselineskip}{7mm}
\begin{titlepage}

\begin{flushright}

{\tt NRCPS-HE-65-2017} \\

\end{flushright}

\vspace{1,5cm}
\begin{center}

{\Large \it  Interaction of Non-Abelian Tensor Gauge Fields

\vspace{0,3cm}

} 

\vspace{1cm}
{\sl  George Savvidy

\bigskip
\centerline{Institute of Nuclear and Particle Physics}
\centerline{ Demokritos National Research Centre}
\centerline{ Ag. Paraskevi,  Athens, Greece}
\bigskip
}

\end{center}

 \begin{abstract}
The non-Abelian tensor gauge fields take values in extended  Poincar\'e 
algebra.  In order to define the invariant Lagrangian we introduce a vector variable in two alternative ways: through the transversal representation 
of the extended Poincar\'e  algebra and through the path integral over the 
auxiliary vector field with the U(1) Abelian action. We demonstrate that path integral formulation allows to fix
the unitary gauge,  derive scattering amplitudes in spinor representation and
to define the gauge invariant  Lagrangian in a curved space-time. 
 
 \end{abstract}

\vspace{1cm}

\end{titlepage}



{\bf 1.}~The concept of local gauge
invariance formulated by Yang and Mills \cite{yang}
allows to define the non-Abelian gauge fields,
to derive the  Lagrangian, the  dynamical field equations and
 to develop a universal point of view on matter interactions  \cite{chern,Weyl:1929fm,Cartan:1923zea}.
It is appealing to extend the gauge principle so that it will define the
interaction of  non-Abelian tensor gauge fields as well\footnote{The research in high spin field theories 
has long and rich history.  One should mention the early works of 
Majorana \cite{majorana},  Fierz \cite{fierz}, Pauli \cite{fierzpauli}, Dirac \cite{dirac},
Ginsburg  and Tamm \cite{ginzburg}, 
 Weinberg \cite{Weinberg:1964cn}, Minkowski \cite{minkowski}, Nambu \cite{Nambu:1968sa},Schwinger \cite{schwinger}, 
Ramond \cite{Ramond:1971gb}, Singh and Hagen \cite{singh}, Fronsdal \cite{fronsdal}, Brink  et.al.   \cite{Bengtsson:1983pd,Bengtsson:1983pg}, 
Berends et.al. \cite{berends},
 Fradkin\cite{fradkin}, Vasiliev \cite{vasiliev}, 
Sagnotti, Sezgin and Sundel \cite{Sagnotti:2005ns}, Metsaev \cite{Metsaev:2005ar}, 
Manvelyan et.al. \cite{Manvelyan:2010jr}  and many other works (see also the references in \cite{Guttenberg:2008qe}). }. 

The recently proposed generalisation  of Yang-Mills theory \cite{Savvidy:2005fi,Savvidy:2005zm,Savvidy:2005ki,Savvidy:2010vb,Savvidy:2010kw,Antoniadis:2011re,Savvidy:2013gsa}
is  based on the extension of the Poincar\'e 
algebra $L_G (\CP )$ with additional generators $L_{ a}^{ \lambda_1 ... \lambda_{n}} $.
 The invariant Lagrangian is defined in terms of a composite gauge field 
 ${\cal A}^{a}_{\mu}(x,e_{\perp})$ and the field strength tensor $\CG^{a}_{\mu\nu}(x,e_{\perp})$
 depending  on the space-time coordinates $x_{\mu}$ and the transversal 
space-like vector variable $e^{\lambda}_{\perp}$  \cite{yukawa1,fierz1,wigner,Savvidy:2010vb,Savvidy:2010kw}.  The  variable 
$e^{\lambda}_{\perp}$ belongs to
irreducible transversal representation $L_{ a}^{ \lambda_1 ... \lambda_{n}} = L_a e^{\lambda_1}_{\perp}...e^{\lambda_n}_{\perp}$ of the algebra $L_G (\CP )$ and fulfils the 
 equations:  
 $e^2_{\perp}=-1$, $(n \cdot e_{\perp})=0$,   $n^2=0$.  These equations
are reminiscent of the Abelian gauge field equations.  We shall elevate this fact into a guiding  principle  postulating that the dynamics  of the vector variables $e_{\lambda}(x)$ is governed by the Abelian $U(1)$ action and the Lagrangian of the tensor gauge fields is  
\beqa\label{lagrangdensity1}
{{\cal L}} &=&   -{1\over 4} \<\CG^{a}_{\mu\nu}(x,e)\CG^{a \mu\nu}(x,e)\rangle~  
= -{1\over 4}\int   \CG^{a}_{\mu\nu}(x,e)\CG^{a \mu\nu}(x,e) ~ e^{ -{i \over 4}\int F^2 (e) d^4x}  ~  \CD e_{\lambda} ~,~~~~~
\eeqa 
where  $F_{\rho\lambda } =  \partial_{\rho} e_{\lambda} - \partial_{\lambda} e_{\rho}$ and 
$\<...\rangle=\int ...\CD e_{\lambda}$ denotes a path integral over $e_{\lambda}(x)$. The advantage of the above formulation   lies in the fact  that the dynamics of the auxiliary variable $e_{\lambda}(x)$ is 
identical to the photon dynamics and therefore naturally describes  the space-like transversal polarisations
$e_{\perp}$  appearing earlier  in  $L_{ a}^{ \lambda_1 ... \lambda_{n}}$ representation
of the $L_G (\CP )$ algebra.  The expansion of the integrand $\CG^{a}_{\mu\nu}(x,e)\CG^{a \mu\nu}(x,e)$  over the auxiliary variable $e^{\lambda}_{\perp}$ expresses the Lagrangian ${{\cal L}}$  in terms of vacuum expectation values 
\be\label{average2}
\CL   = \sum^{\infty}_{s=0}~{1\over s!}
\CL_{\lambda_1 ... \lambda_{s}} ~ \<e^{\lambda_1}_{\perp}...e^{\lambda_s}_{\perp}\rangle~. 
\ee
 The $L_G (\CP )$ generators have the form $L_{ a}^{ \lambda_1 ... \lambda_{n}} = L_a e^{\lambda_1}_{\perp}...e^{\lambda_n}_{\perp}$
and the  vacuum expectation values $ \<L_{ a}^{ \lambda_1 ... \lambda_{n}} \vert L_{ b}^{ \lambda_{n+1} ... \lambda_{s}}  \rangle = \delta_{ab}~ \<e^{\lambda_1}_{\perp}...e^{\lambda_n}_{\perp}   e^{\lambda_{n+1}}_{\perp}...e^{\lambda_s}_{\perp}\rangle $  play  the role of the Killing metric. The Killing metric contracts only  the space-like components and guarantee 
the absence of negative norm time-like components of the tensor gauge fields 
in the Lagrangian $\CL$.

{\bf 2.}~The non-Abelian tensor gauge fields 
$
A^{a}_{\mu\lambda_1 ... \lambda_{s}}(x),~s=0,1,2,...
$
are totally symmetric in indices  $  \lambda_1 ... \lambda_{s}  $ and 
are the components  of the field ${\cal A}_{\mu}(x,e)$ \cite{Savvidy:2005fi,Savvidy:2005zm,Savvidy:2005ki}:
\be\label{gaugefield1}
{\cal A}_{\mu}(x,e)=\sum_{s=0}^{\infty} {1\over s!} ~A^{a}_{\mu\lambda_{1}...
\lambda_{s}}(x)~L_{a}e^{\lambda_{1}}...e^{\lambda_{s}}
\ee
taking values in the extended  Poincar\'e 
algebra. The $L_G (\CP )$  algebra   \cite{Savvidy:2005ki,Savvidy:2010vb,Savvidy:2010kw,Antoniadis:2011re} is:
\beqa\label{extensionofpoincarealgebra}
~&&[P^{\mu},~P^{\nu}]=0,\nn\\
~&&[M^{\mu\nu},~P^{\lambda}] = \eta^{\nu\lambda }~P^{\mu}
- \eta^{\mu\lambda  }~P^{\nu} ,\nn\\
~&&[M^{\mu \nu}, ~ M^{\lambda \rho}] = \eta^{\mu \rho}~M^{\nu \lambda}
-\eta^{\mu \lambda}~M^{\nu \rho} +
\eta^{\nu \lambda}~M^{\mu \rho}  -
\eta^{\nu \rho}~M^{\mu \lambda} ,\nonumber\\
~&&[P^{\mu},~L_{a}^{\lambda_1 ... \lambda_{s}}]=0, \nn\\
~&&[M^{\mu \nu}, ~ L_{a}^{\lambda_1 ... \lambda_{s}}] =
\eta^{\nu \lambda_1}~L_{a}^{\mu \lambda_2... \lambda_{s}}
-\eta^{\mu\lambda_1}~L_{a}^{\nu\lambda_2... \lambda_{s}}
+...+
\eta^{\nu \lambda_s}~ L_{a}^{\mu \lambda_1... \lambda_{s-1}} -
\eta^{\mu \lambda_s}~L_{a}^{\nu \lambda_1... \lambda_{s-1}} ,\nonumber\\
~&&[L_{a}^{\lambda_1 ... \lambda_{k}}, L_{b}^{\lambda_{k+1} ... \lambda_{s}}]=if_{abc}
L_{c}^{\lambda_1 ... \lambda_{s}} .
\eeqa
It  incorporates the Poincar\'e    and internal algebra $L_G$ with structure constants $f_{abc}$.  The  generators
$L_{a}^{\lambda_1 ... \lambda_{s}}$  carry the {\it  internal charges  and spins}.
 There is no conflict with the Coleman-Mandula theorem
\cite{Coleman:1967ad} because
the theorem applies to the symmetries that act on S-matrix elements and 
not to symmetries which acts on the gauge field
$\CA_{\mu}(x,e)$.
The algebra (\ref{extensionofpoincarealgebra}) is invariant  with respect to the  "gauge"
transformations \cite{Savvidy:2010vb,Savvidy:2010kw,Antoniadis:2011re}:
\beqa\label{isomorfism}
& L_{a}^{\lambda_1 ... \lambda_{s}} \rightarrow L_{a}^{\lambda_1 ... \lambda_{s}}
+ \sum_{1} \chi~ P^{\lambda_1}L_{a}^{\lambda_2 ... \lambda_{s}}+
\sum_{2} \chi ~P^{\lambda_1} P^{\lambda_2} L_{a}^{\lambda_3 ... \lambda_{s}} +...+
\chi^s ~P^{\lambda_1}... P^{\lambda_s} L_{a} \nn\\
& M^{\mu\nu} \rightarrow M^{\mu\nu},~~~~
P^{\lambda} \rightarrow P^{\lambda},
\eeqa
where the sums  $\sum_{1},\sum_{2},... $ are over all inequivalent index permutations
and $\chi$ is a scalar. The invariance retains if one consider any translationally invariant vector $n^{\lambda}$ instead of $P^{\lambda}$. 
 The representations of $L_G (\CP )$ are therefore defined modulo longitudinal terms proportional to the momentum $P^{\lambda}$.
The algebra $L_G(\CP)$ has representation in terms of the differential operators:
\beqa\label{represofextenpoincarealgebra}
~&& P^{\mu} = k^{\mu} ,\nn\\
~&& M^{\mu\nu} = i(k^{\mu}~ {\partial\over \partial k_{\nu}}
- k^{\nu }~ {\partial \over \partial k_{\mu}}) + i(e^{\mu}~ {\partial\over \partial e_{\nu}}
- e^{\nu }~ {\partial \over \partial e_{\mu}}),\nn\\
~&& L_{a}^{\lambda_1 ... \lambda_{s}} =e^{\lambda_1}...e^{\lambda_s} \otimes L_a~.
\eeqa
The irreducible transversal representation
is defined by invariant equations \cite{yukawa1,fierz1,wigner}:
\be\label{constraint}
n^2=0,~~~(n \cdot  e_{\perp})=0,~~~e^2_{\perp}=-1,
\ee
where $n_{\lambda}$ is an arbitrary light-like vector\footnote{A vector variable, 
in addition to the space-time coordinate $x$,  was introduced
earlier by Yukawa \cite{yukawa1}, Fierz \cite{fierz1} and Wigner\cite{wigner}.    
}.  
These equations have the solution 
\be\label{solution}
e^{\mu}_{\perp}=  \chi n^{\mu} + e^{i \varphi} e^{\mu}_{+}
+e^{-i \varphi} e^{\mu}_{-} ,
\ee
where $e^{\mu}_{\pm}$ are helicity polarisation 
vectors,   the variables   $\chi$ and  $\varphi$ are defined 
on the cylinder $ \varphi \in S^1, \chi \in R^1 $.  The 
 representation of  the generators
$L_{a}^{ \lambda_1 ... \lambda_{s}} $ therefore is:
\be\label{trasversalgenera11}
L_{a}^{\bot~ \lambda_1 ... \lambda_{s}}= \prod^{s}_{n=1} ( \chi n^{\lambda_n} + e^{i \varphi} e^{\lambda_n}_{+}
+e^{-i \varphi} e^{\lambda_n}_{-}) \otimes L_a~
\ee
and it is  transversal  because of (\ref{constraint}):
\be\label{transversality}
n_{\lambda_1}L_{a}^{\bot \lambda_1 ... \lambda_{s}}=0,~~~~s=1,2,...
\ee
Opening the brackets 
in (\ref{trasversalgenera11}) and separating  terms containing the
 variables $n^{\lambda_i}$ we shall get  the first term of the form
$\prod^{s}_{n=1} (e^{i \varphi} e^{\lambda_n}_{+}
+e^{-i \varphi} e^{\lambda_n}_{-})$ which represents the {\it helicity generators }
$(L^{\lambda_1 ... \lambda_{s}}_{a +\cdot\cdot\cdot+},...,L^{\lambda_1 ... \lambda_{s}}_{a -\cdot\cdot\cdot-})$ with helicity spectrum 
$
h=(+s, +s-2,......, -s+2, -s)
$.  The rest of the terms  contain  $n^{\lambda}$ and correspond to  the transformation (\ref{isomorfism}) of $L_{a}^{\bot~ \lambda_1 ... \lambda_{s}}$.

Substituting the representation (\ref{trasversalgenera11}) of the
 $L_{a}^{\bot \lambda_1 ... \lambda_{s}}$ into the
expansion (\ref{gaugefield1}) and collecting terms in front of the helicity generators
$(L^{\lambda_1 ... \lambda_{s}}_{a +\cdot\cdot\cdot+},...,L^{\lambda_1 ... \lambda_{s}}_{a -\cdot\cdot\cdot-})$  we shall get (see Appendix for details)
\beqa\label{gaugefield2}
{\cal A}_{\mu}(x,e)&=&   \sum_{s=0}^{\infty} {1\over s!} ~
 (~ A_{\mu +\cdot\cdot\cdot+}~ +...+~
 A_{\mu -\cdot\cdot\cdot-}~  ),
\eeqa
where $s$ is the number of helicity indices and 
$
A_{\mu +\cdot\cdot\cdot-} =A^{a}_{\mu\lambda_{1}...
\lambda_{s}}~L^{\lambda_1 ... \lambda_{s}}_{a +\cdot\cdot\cdot-}.
$ 
It follows from  (\ref{gaugefield2})  that the time components $\lambda_i=0,~i=1,...,s$ of the tensor gauge fields have been  
gauged away.  The rest of the time components 
$(A_{0 +\cdot\cdot\cdot+},...,A_{0 -\cdot\cdot\cdot-})$,  when the first index $\mu =0$
is equal to zero,  will be gauged away by the  gauge transformation   \cite{yang,Savvidy:2005ki,Savvidy:2010vb}:
\be\label{extendedgaugetransformation}
\CA^{'}_{\mu}(x,e) = U(\xi)  \CA_{\mu}(x,e) U^{-1}(\xi) -{i\over g}
\partial_{\mu}U(\xi) ~U^{-1}(\xi),
\ee
where $U(\xi)=\exp{(i \xi(x,e))}$ and the parameter  $\xi(x,e)$
depends on the vector variable   \cite{Savvidy:2005ki,Savvidy:2010vb}:
$$
\xi(x,e)=  \sum_s {1\over s!}~\xi^{a}_{\lambda_1 ... \lambda_{s}}(x) ~L_{a}^{\lambda_1 ... \lambda_{s}} .
$$
The components $\xi^{a}_{\lambda_1 ... \lambda_{s}}(x)$ are totally symmetric.
The commutator of covariant derivatives
$
\nabla^{ab}_{\mu} = (\partial_{\mu}-ig \CA_{\mu}(x,e))^{ab}
$
defines the field strength tensor
\beqa\label{fieldstrengthgeneral}
\CG_{\mu\nu}(x,e)&=& \partial_{\mu} \CA_{\nu}(x,e) - \partial_{\nu} \CA_{\mu}(x,e) -
i g [ \CA_{\mu}(x,e)~\CA_{\nu}(x,e)] \nn\\
& =& \sum_s {1\over s!}~G^{a}_{\mu\nu, \lambda_1 ... \lambda_{s}}(x) ~L_{a}^{\lambda_1 ... \lambda_{s}},
\eeqa
which transforms homogeneously 
$
\CG^{'}_{\mu\nu}(x,e)) = U(\xi)  \CG_{\mu\nu}(x,e) U^{-1}(\xi)
$
and the gauge invariant Lagrangian is  defined as \cite{Savvidy:2005fi,Savvidy:2005zm,Savvidy:2005ki}
\be\label{lagrangdensity}
{{\cal L}}(x)= \<{{\cal L}}(x,e)\rangle =  -{1\over 4} \<\CG^{a}_{\mu\nu}(x,e)\CG^{a \mu\nu}(x,e)\rangle.
\ee
Using the  $L_G(\CP)$ generators  
(\ref{trasversalgenera11}) one can calculate 
 the Killing metric \cite{Savvidy:2010vb,Savvidy:2010kw,Savvidy:2013gsa}:
\beqa\label{killingform}
L_G:~~~~~~&&~~~ \<L_{a}\vert L_{b} \rangle  =\delta_{ab}, \label{finale0}\\
L_{\CP}:~~~~~~~&& ~~~\<P^{\mu} \vert P^{\nu }  \rangle ~=0\nn\\
&&~~~\<M_{\mu\nu} \vert P_{\lambda  }  \rangle ~  =0\label{poincare0}\nn\\
&& ~~~\<M^{\mu\nu} \vert M^{\lambda \rho }  \rangle =\eta^{\mu\lambda } \eta^{\nu\rho}
-\eta^{\mu\rho} \eta^{\nu \lambda }\nn\\
L_G(\CP):~~~~~&&~~~~\<P^{\mu}\vert L_{a}^{\bot~ \lambda_1 ... \lambda_{s}}\rangle  =0,\nn\\
&&~~~~\<M^{\mu\nu}\vert L_{a}^{\bot~ \lambda_1 ... \lambda_{s}}\rangle =0,\nn\\
&&~~~~\<L^{\bot~\lambda_1...\lambda_n}_{a} \vert L^{\bot~\lambda_{n+1}....\lambda_{2s+1}}_{b} \rangle  = 0,~~~~~~~~~s=0,1,2,3,...\nn\\
&&~~~~\<L^{\bot~\lambda_1...\lambda_n}_{a} \vert L^{\bot~\lambda_{n+1}....\lambda_{2s}}_{b} \rangle  =
\delta_{ab}~ s!~(\bar{\eta}^{\lambda_1 \lambda_2}  \bar{\eta}^{\lambda_3 \lambda_4}...
\bar{\eta}^{\lambda_{2s-1} \lambda_{2s}} +\textrm{perm}),\nn
\eeqa
where $\bar{\eta}^{\lambda_1\lambda_2}$ is the projector  into
the two-dimensional plane transversal to the light-like vector  $n^{\lambda}$ \cite{schwinger}:
\be\label{progector}
\bar{\eta}^{\lambda_1\lambda_2} =  - \eta^{\lambda_1 \lambda_2} + { n^{\lambda_1}\bar{n}^{\lambda_2}
+\bar{n}^{\lambda_1}n^{\lambda_2} \over n \bar{n}}=
\sum_{i=\pm} e^{ \lambda_1}_i  e^{ \lambda_2}_i,~~~~~~~
n_{\lambda_1 }\bar{\eta}^{\lambda_1\lambda_2}=
n_{\lambda_2 }\bar{\eta}^{\lambda_1\lambda_2}=0,
\ee
where $n=(n_0, \vec{n})$ and  $\bar{n}^{\mu}=(n_0,-\vec{n})$
and  the Killing metric is transversal (\ref{constraint}):
\be\label{transversality1}
n_{\lambda_i} \<L^{\bot~\lambda_1...\lambda_n}_{a}; L^{\bot~\lambda_{n+1}....
\lambda_{2s}}_{b} \rangle =0,~~~~i=1,2,...2s.
\ee 
Expanding the   ${{\cal L}}(x,e)$ in (\ref{lagrangdensity}) over the vector variable $e^{\lambda}$ we shall get 
 \cite{Savvidy:2005fi,Savvidy:2005zm,Savvidy:2005ki}:
\be\label{average1}
\CL(x) = \<\CL(x,e)\rangle = \sum^{\infty}_{s=0}~{1\over s!}
\CL_{\lambda_1 ... \lambda_{s}}(x) ~ \<e^{\lambda_1}...e^{\lambda_s}\rangle~.
\ee
In particular, the term quadratic in $e^{\lambda}$ is
\be
{{\cal L}}_2 = \CL_{\lambda_1\lambda_2} \<e^{\lambda_1}e^{\lambda_2}\rangle= -{1\over 4}(G^{a}_{\mu\nu,\lambda_1}G^{a \mu\nu,}_{~~~~\lambda_2}
+ G^{a}_{\mu\nu,\lambda_1\lambda_2} G^{a \mu\nu})\bar{\eta}^{\lambda_1\lambda_2} 
\ee
or  using the bar symbol  over the indices  indicating
the transversal Killing metric  contraction   (\ref{progector}) we get
\be\label{L2lagrangian}
{{\cal L}}_2 = -{1\over 4}  G^{a}_{\mu\nu,\bar{\lambda}} ~ G^{a \mu\nu,\bar{\lambda}}
 -{1\over 4}   G^{a~~~~\bar{\lambda}}_{ \mu\nu,\bar{\lambda }} ~G^{a \mu\nu}   .
\ee

 {\bf 3.}~In the  
transversal representation (\ref{trasversalgenera11}) of the $L_G (\CP )$  algebra the tensor gauge fields (\ref{gaugefield2}) are partially gauged  and the residual gauge transformation of (\ref{extendedgaugetransformation}), (\ref{polygauge1}) takes the 
following  form
 \cite{Savvidy:2005fi,Savvidy:2005zm,Savvidy:2005ki}
 \beqa\label{polygauge2} 
 A^{'}_{\mu} &=& A_{\mu} +  \nabla_{\mu}\xi ,~~~~~\\
 A^{'}_{\mu \pm} &=&  A_{\mu \pm}  + \partial_{\mu} \xi_{\pm} -
i g [A_{\mu}, \xi_{\pm}] -i g [A_{\mu\pm}, \xi],\nn\\
A^{'}_{\mu \pm\pm}& =&  A_{\mu\pm\pm} +  \nabla_{\mu} \xi_{\pm\pm} 
-i g (  [A_{\mu  \pm}, \xi_{\pm }] +
[A_{\mu \pm }, \xi_{ \pm}]+
[A_{\mu\pm\pm}, \xi]),\nn\\
A^{'}_{\mu +-}& =&  A_{\mu+-} +  \nabla_{\mu} \xi_{+-} 
-i g (  [A_{\mu  +}, \xi_{- }] +
[A_{\mu - }, \xi_{ +}]+
[A_{\mu+-}, \xi]),\nn\\
.........&.&............................ \nn
\eeqa
where 
$
A_{\mu \pm}  = A^{a}_{\mu \lambda} L^{\perp \lambda}_{a \pm} ,~~
\xi_{\pm} = \xi^{b}_{\lambda} L^{\perp \lambda}_{b \pm},~~
A_{\mu \pm \pm }  = A^{a}_{\mu  \nu \lambda} L^{\perp \nu \lambda}_{a \pm \pm} ,~~
\xi_{\pm\pm} = \xi^{b}_{\nu \lambda} L^{\perp \nu \lambda}_{b \pm\pm}
$
and so on. Using the residual gauge transformation (\ref{polygauge2}) we  
can impose the 
Lorentz gauge on the rest of the field components: 
\be\label{gaugefix}
 \partial^{\mu}  A^{'}_{\mu +....-}=0.
\ee
Indeed, we shall impose the standard Lorentz gauge $\partial^{\mu}  A^{'}_{\mu }=0$    by solving the equation  
\be\label{gaugefixing1}
\partial^{\mu}  A_{\mu } +    \partial^{\mu} \nabla_{\mu} \xi=0,~~~\xi = -(\partial^{\mu} \nabla_{\mu})^{-1} \partial^{\nu}  A_{\nu }~.
\ee
Imposing a similar  gauge $\partial^{\mu}  A^{'}_{\mu \pm}=0$ on the tensor gauge field  we get 
\be\label{gaugefixing2}
\partial^{\mu} A_{\mu \pm} +    \partial^{\mu} \nabla_{\mu} \xi_{\pm}  -i g \partial^{\mu}  [A_{\mu\pm}, \xi]=0,
\ee
which can be solved by using the gauge parameters $\xi_{\pm}$
\be
\xi_{\pm}  = -(\partial^{\mu} \nabla_{\mu} )^{-1} 
\partial^{\nu} ( A_{\nu \pm}   -i g   [A_{\nu\pm}, \xi]),
\ee
The gauge condition $\partial^{\mu} A^{'}_{\mu \pm \pm} =0$ on 
the rank-3 gauge field can also be resolved by 
\beqa\label{gaugefixing3}
\xi_{\pm\pm}  &=& -(\partial^{\mu} \nabla_{\mu} )^{-1} \partial^{\nu} 
( A_{\nu \pm\pm}   -2 i g   [A_{\nu\pm}, \xi_{\pm}] -i g [A_{\nu\pm\pm}, \xi]),\nn\\
\xi_{+-}  &=& -(\partial^{\mu} \nabla_{\mu} )^{-1} \partial^{\nu} 
( A_{\nu +-}   - i g   [A_{\nu+}, \xi_{-}]  - i g   [A_{\nu-}, \xi_{+}]  -i g [A_{\nu+-}, \xi]).
\eeqa
Thus the two-stage gauge fixing procedure  (\ref{constraint}) - 
(\ref{gaugefield2})  and (\ref{gaugefix}) - (\ref{gaugefixing3})
\be\label{gaugefixing0}
A_{\mu +\cdot\cdot\cdot-} =A^{a}_{\mu\lambda_{1}...
\lambda_{s}}~L^{\lambda_1 ... \lambda_{s}}_{a +\cdot\cdot\cdot-} ~~~~~~\text{and}~~~~~~
 \partial^{\mu}  A^{'}_{\mu +....-}=0
\ee
allows to completely exclude the time-like 
components of the tensor gauge fields leaving only pure transversal helicity 
polarisations.

 {\bf 4.}~As it was  demonstrated above, the vector variable $e^{\lambda}$ has the 
properties which are reminiscent of the Abelian gauge field.  We shall elevate this fact into a guiding principle  postulating that the 
dynamics  of the vector variables $e^{\lambda}(x)$ is governed by the Abelian $U(1)$ action.
We shall consider now the Killing metric (\ref{killingform}) as a correlation 
function of the Abelian $U(1)$ gauge field $e^{\lambda}(x)$ defined in terms of the Feynman path integral: 
\beqa\label{Faddeev}
 \< Te^{\lambda_1}(x_1)...e^{\lambda_s}(x_s) \rangle = \int   e^{i \int (-{1\over 4}F_{\alpha\beta}(e)F^{\alpha\beta}(e) +J_{\alpha} e^{\alpha} ) } e^{\lambda_1}(x_1)...e^{\lambda_s}(x_s)  \CD e_{\lambda}(x) ,
\eeqa 
where the external current  is conserved:
$
 \partial^{\alpha}J_{\alpha} =0
 $ 
 \cite{schwinger}.
The action for the vector variable $e^{\lambda}(x)$ in the axial gauge has the form 
\beqa
   \int \{-{1\over 4}F_{\alpha\beta}(e)F^{\alpha\beta}(e)
 + \lambda^2  e^{\alpha}(x) n_{\alpha}n_{\beta}  e^{\beta}(x)
\}   d^4x
 = {1\over 2} \int e^{\alpha}(x) H_{\alpha\beta} e^{\beta}(x) 
 d^4x ,~~~
\eeqa
where
$
H_{\alpha\beta}=  \eta_{\alpha\beta} \partial^2   
-   \partial_{\alpha} \partial_{\beta}    + \lambda^2  n_{\alpha} n_{\beta},~~~ 
H_{\alpha\beta}=  -\eta_{\alpha\beta} k^2  
+ k_{\alpha}     k_{\beta}   + \lambda^2  n_{\alpha} n_{\beta}~.
$
The vector variable propagator 
$
H_{\alpha\beta}(k) \Delta^{\beta\gamma}(k) =i \delta_{\alpha}^{\gamma}
$
has the following expression:
\beqa
&&  \Delta^{\alpha\beta}(k)=  \Big( -\eta^{\alpha\beta}  + {k^2  -\lambda^2 n^2 \over 
\lambda^2 (k \cdot n)^2  } k^{\alpha} k^{\beta}   +{ k^{\alpha} n^{\beta} +n^{\alpha} k^{\beta}  \over 
  (k \cdot n)  } \Big) {i \over  k^2  +i \epsilon } .\nn
\eeqa
In the limit  $\lambda^2 \rightarrow \infty$
\beqa
&&\Delta^{\alpha\beta}(k)=   
  i  \Big(-\eta^{\alpha\beta}   
- {  n^2             \over 
  (k \cdot n)^2   }  k^{\alpha} k^{\beta}    +{ k^{\alpha} n^{\beta} +n^{\alpha} k^{\beta}  \over 
  (k \cdot n)  }   \Big)  {i  \over k^2   +i \epsilon}  \nn
\eeqa
the propagator is explicitly transversal 
\be\label{trans}
n_{\alpha}\Delta^{\alpha\beta}(k)=0.
\ee
In the light-cone gauge, when
$
n^2 = 0, 
$
we shall get 
\beqa
&\Delta^{\alpha\beta}(x-y) =  \< 0\vert T e^{\alpha}(x) e^{\beta}(y) \vert 0\rangle 
=   \int {d^d k \over (2\pi)^d}    \Big(-\eta^{\alpha\beta}   
   +{ k^{\alpha} n^{\beta} +n^{\alpha} k^{\beta}  \over 
  (k \cdot n)  }   \Big)  {  i \over k^2   +i \epsilon} e^{-i k\cdot (x-y) } ~~~~~~~~
\eeqa
and  the correlation function at the coincident points $x \rightarrow y$
is:
\beqa
&&\< 0\vert T e^{\alpha}(x) e^{\beta}(x) \vert 0\rangle 
=  \int {d^d k \over (2\pi)^d}   \Big(-\eta^{\alpha\beta}   
  +{ k^{\alpha} n^{\beta} +n^{\alpha} k^{\beta}  \over 
  (k \cdot n)  }   \Big) {i \over k^2   +i \epsilon}  . \nn
\eeqa
The above integral depends on the covariant expression of the form
\be
 \< 0\vert T e^{\alpha}(x) e^{\beta}(x) \vert 0\rangle = -A \eta^{\alpha\beta}   
 +B( n^{\alpha} \bar{n}^{\beta} + \bar{n}^{\alpha} n^{\beta}),
\ee
where $n^{\alpha} =(n_0,\vec{n})$, $\bar{n}^{\alpha} =(n_0,- \vec{n}) $ and 
$n^2=\bar{n}^2=n^2_0 -\vec{n}^2=0$.   Calculating 
the trace $I$ and contraction with $n_{\alpha}$, we shall get 
two equations:
$
-4 A    + 2 B (n \cdot \bar{n})=I,~-A +  B (n \cdot \bar{n}) =0, 
$
therefore  $B=A /(n \cdot \bar{n})$ and  
\beqa
A=   \int {d^d k \over (2\pi)^d}      {i \over k^2  +i \epsilon}~. 
\eeqa
 These lead us to the expression for the vector variable correlation function 
\be\label{matrix}
{1\over A }  \< 0\vert T e^{\alpha}(x) e^{\beta}(x) \vert 0\rangle  =  -  \eta^{\alpha\beta}  
  + { n^{\alpha} \bar{n}^{\beta} + \bar{n}^{\alpha} n^{\beta} \over
 (n \cdot \bar{n}) } ,
\ee
which is transversal to the vector $n_{\alpha}$ (\ref{trans}). 
Because $n^2=0$,  the matrix (\ref{matrix})
has two eigenvectors $e_{\pm}$ in the space-like directions: 
\be
n^2 =\bar{n}^2=0,~~~(n \cdot e_{ \pm})=0~~~(\bar{n} \cdot e_{ \pm})=0,~~~e^2_{ \pm}=0,
~~~~ ( e_+ \cdot e_- ) =1 
\ee  
and  
\be\label{killing}
\bar{  \eta}^{\alpha\beta} ={1\over A } \< 0\vert T e^{\alpha}(x) e^{\beta}(x) \vert 0\rangle  =   -\eta^{\alpha\beta}  
+ { n^{\alpha} \bar{n}^{\beta} + \bar{n}^{\alpha} n^{\beta} \over
 (n \cdot \bar{n}) }  =
\sum_{i=\pm} e^{ \alpha}_i  e^{  \beta}_i,
\ee
Using the Wick theorem  we shall get 
the metric   (\ref{killingform}).   Thus the Killing metric, expressed  in terms of the  Feynman path integral (\ref{Faddeev}) and in terms of traces of the $L_G(\CP)$ generators  (\ref{killingform}) in the transversal representation  (\ref{trasversalgenera11}) coincide.

The conclusion is that the Lagrangian (\ref{lagrangdensity1}), (\ref{lagrangdensity}) for the tensor gauge fields 
can be defined in {\it flat space-time} in two equivalent ways: either by using transversal representation of the $L_G(\CP)$ generators  (\ref{killingform}) or by postulating that the dynamics  of the vector variable 
is governed by the Abelian action (\ref{Faddeev}). In both cases the vector variable has two transversal 
polarisations and the corresponding Killing metric contracts transversal 
space-like components  of the tensor gauge fields. 
 Being equivalent in flat space-time they are not equivalent in curved. 
The advantage of the path integral formulation lies  in the fact that it opens a prospect to define the gauge invariant  
Lagrangian in {\it curved space-time}. 

{\bf 4.}~
The Lagrangian (\ref{lagrangdensity1}), (\ref{lagrangdensity}) defines the propagation 
of free tensor gauge bosons and their interactions.   
The high-spin bosons  interact
through the triple and quartic interaction vertices and a dimensionless coupling 
constant. In order to calculate scattering amplitudes it is convenient to use 
spinor representation
of amplitudes developed in \cite{Parke:1986gb,Dixon:1996wi,
Witten:2003nn,Britto:2005fq}.
A scattering amplitude $
M_n=M_n(\lambda_1,\tilde{\lambda}_1,h_1;~...;~\lambda_n,\tilde{\lambda}_n,h_n)
$
for the
massless particles of momenta $p_i$ and polarisation tensors $\varepsilon_i$ ~$(i=1,...,n)$  can be represented in terms of spinors:
\be\label{spinors12}
k_{a\dot{a}}= \lambda_a \tilde{\lambda}_{\dot{a}},~~e^{+}_{a\dot{a}}
={\mu_a \tilde{\lambda}_{\dot{a}} \over <\mu,\lambda>},~~e^{-}_{a\dot{a}}
={\lambda_a \tilde{\mu}_{\dot{a}} \over [\lambda,\mu]},
\ee
where
$$
\lambda_a = (\sqrt{k^+},{k_x +i k_y \over \sqrt{k^+}}),~~~
\tilde{\lambda}_{\dot{a}} = (\sqrt{k^+},{k_x - i k_y \over \sqrt{k^+}}),~~~~k^+=k_t + k_z~,
$$
and $\mu_a$ is a reference spinor.  In Yang-Mills theory the interaction vertex $VVV$
is
\be\label{VVV}
{{\cal V}}^{abc}_{\alpha\beta\gamma}(k,p,q)= -i g f^{abc}
 F_{\alpha\beta\gamma}(k,p,q) =
-i g f^{abc} [\eta_{\alpha\beta} (p-k)_{\gamma}+ \eta_{\alpha\gamma} (k-q)_{\beta}
 + \eta_{\beta\gamma} (q-p)_{\alpha}] .
\ee
In (\ref{lagrangdensity1}), (\ref{lagrangdensity})  the interaction vertex $VTT$, the  tensor-vector-tensor vertex is \cite{Savvidy:2005ki}
\be\label{VTT}
 \CV^{abc}_{\alpha\acute{\alpha}\beta\gamma\acute{\gamma}}(k,p,q)
=-  i g f^{abc} [\eta_{\alpha\beta} (p-k)_{\gamma}+ \eta_{\alpha\gamma} (k-q)_{\beta}
 + \eta_{\beta\gamma} (q-p)_{\alpha}] \bar{\eta}_{\acute{\alpha}\acute{\gamma}}~,
\ee
where the indices $(a, \alpha \acute{\alpha},k)$ belong to the tensor
gauge boson, $(b, \beta,p)$ to the vector gauge boson and $(c, \gamma \acute{\gamma},q)$
to the second tensor gauge boson.   In massless theory the momentum conservation
$\delta(k + p +q)$ equation has the solution
$
k  = (\omega, 0, 0, r),~~p  =
(\omega, 0, 0, r),~~q = (-2\omega, 0, 0, -2r)
$
$(\omega^2 = r^2)$ that can be deformed by a complex parameter $z$
\cite{ Dixon:1996wi,Parke:1986gb,
Witten:2003nn,Britto:2005fq,Benincasa:2007xk},
$
k  = (\omega, z, iz, r),~~p  =
(\omega, -z, -iz, r),~~q = (-2\omega, 0, 0, -2r)
$
and the corresponding polarisation vectors
$$
e^{+}_{k}={1\over \sqrt{2}} ({z \over \omega}, 1, -i, -{z \over r} ),~~
e^{+}_{p}={1\over \sqrt{2}}(-{z \over \omega}, 1, -i, {z \over r} ),~~
e^{-}_{q}={1\over \sqrt{2}}(0, 1,  i ,0 )
$$
are orthogonal to the momenta
$
k \cdot e^{+}_{k} =0,~~p \cdot e^{+}_{p} =0,~~q \cdot e^{-}_{q} =0.
$
Computing the matrix element of the YM vertex $VVV$ (\ref{VVV}) we shall get:
\beqa\label{vvv}
&M_{YM}(+1,+1,-1)= F_{\alpha\beta\gamma}(k,p,q)~ e^{+\alpha}_k e^{+\beta}_p e^{-\gamma}_q=\nn\\
&= (e^+_{k} \cdot e^+_{p})~ (p-k \cdot e^-_{q}) +
 (e^+_{p} \cdot e^-_{q})~ (q-p \cdot e^+_{k})+
 (e^-_{q}  \cdot  e^+_{k})~ (k-q \cdot e^+_{p})=\nn\\
&= 2 (e^+_{k} \cdot e^+_{p})~ (p \cdot e^-_{q}) +
 2 (e^+_{p} \cdot e^-_{q})~ (q \cdot e^+_{k})+
 2 (e^-_{q}  \cdot  e^+_{k})~ (k \cdot e^+_{p}).
\eeqa
In a spinor representation (\ref{spinors12}) the amplitude (\ref{vvv}) is
$$
M_{YM}(+1,+1,-1)=   {[1,2]^4 \over [1,2] [2,3] [3,1]},
$$
where the reference spinors have been chosen as $\mu^{(k)} = \lambda^{(p)}$, 
$\mu^{(p)} = \lambda^{(q)}$  and  $\mu^{(q)} = \lambda^{(p)}$. 
For the   $VTT$  amplitude
$
M_{GYM}= \varepsilon^{\alpha\acute{\alpha}}(k) e^{\beta}(p)
\varepsilon^{\gamma\gamma^{'}}(q)~
\CV_{\alpha\acute{\alpha}\beta\gamma\acute{\gamma}}(k,p,q)~
\delta(k + p +q)
$
 we shall get 
\be
M_{GYM}(+2,+1,-2)= 2 \Big(e^+_{k} \cdot e^+_{p})~ (p \cdot e^-_{q}) +
  (e^+_{p} \cdot e^-_{q})~ (q \cdot e^+_{k})+
  (e^-_{q}  \cdot  e^+_{k})~ (k \cdot e^+_{p}) \Big) (e^+_{k} \cdot e^-_{q}).
\ee
The last term has the following form: 
\be
 (e^+_{k} \cdot e^-_{q}) =  {<\mu^{(k)}, \lambda^{(q)}>  [ \lambda^{(k)}, \mu^{(q)}] \over 
 <\mu^{(k)}, \lambda^{(k)}>  [ \lambda^{(q)} ,\mu^{(q)}] } = 
 \Big( {[1,2] \over  [2,3]} \Big)^2~~,
\ee
and in spinor representation we shall get
\be
M_{GYM}(+2,+1,-2)=  {[1,2]^4 \over [1,2] [2,3] [3,1]} \Big( {[1,2] \over  [2,3]} \Big)^2~.
\ee
The reference spinors for both tensor bosons are equal  to each other: 
$\mu^{(k)} = \mu^{(q)}$.
Considering the high spin tensors one can find
\be\label{GYMvertex}
M_{GYM}(+s,+1,-s)=  {[1,2]^4 \over [1,2] [2,3] [3,1]} \Big( {[1,2] \over  [2,3]} \Big)^{2s-2}~,
\ee
which reduces to the YM  amplitude  when $s=1$. 
The TTT-amplitude for particles of helicities $(h_1,h_2,h_3)$   has the following
general form \cite{Bengtsson:1983pd,Bengtsson:1983pg,Benincasa:2007xk,Georgiou:2010mf}:
\beqa\label{dimensionone1}
M_3 &=& g f^{abc} <1,2>^{-2h_1 -2h_2 -1} <2,3>^{2h_1 +1} <3,1>^{2h_2 +1},~~~~h_3= -1 - h_1 -h_2, \nn\\
M_3 &=& g f^{abc} [1,2]^{2h_1 +2h_2 -1} [2,3]^{-2h_1 +1} [3,1]^{-2h_2 +1},~~~~~h_3= 1 - h_1 -h_2.
\eeqa
In particular, considering the interaction between bosons of helicities $(+s,+ 1,-s)$  
 one can get convinced that the general expression (\ref{dimensionone1})   is
in full agreement with the direct calculation (\ref{GYMvertex}).

 {\bf 5.}~  The advantage of the path integral formulation of the Lagrangian (\ref{lagrangdensity1}), (\ref{lagrangdensity}) is that it can be defined in a background gravitational field as well
 \cite{Savvidy:2005fi,Savvidy:2005zm,Savvidy:2005ki}:
\beqa\label{Faddeev1}
{{\cal L}} = -{1\over 4}\int  \CG^{a}_{\mu\nu}(x,e)\CG^{a \mu\nu}(x,e)  e^{ - i{1\over 4}\int F_{\alpha\beta}(e)F^{\alpha\beta}(e) \sqrt{-g} d^4x}  \CD e_{\lambda}(x) ,
\eeqa 
where the field strength tensors are well defined in terms of ordinary derivatives: 
\beqa
F_{\alpha\beta}(e) &=&   e_{\beta;\alpha} -  e_{\alpha;\beta}=
  \partial_{\alpha} e_{\beta} - \partial_{\beta} e_{\alpha}\nn\\
  \CG_{\mu\nu}(x,e)&=&  \CA_{\nu;\mu}(x,e) - \CA_{\mu;\nu}(x,e) -
i g [ \CA_{\mu}(x,e)~\CA_{\nu}(x,e)] \nn\\
&=& \partial_{\mu} \CA_{\nu}(x,e) - \partial_{\nu} \CA_{\mu}(x,e) -
i g [ \CA_{\mu}(x,e)~\CA_{\nu}(x,e)] .
\eeqa
It follows therefore that the extended gauge invariance,  
(\ref{extendedgaugetransformation}) and  (\ref{polygauge1}), still holds.

{\bf 6.}~ We calculated the scattering amplitudes of non-Abelian tensor gauge bosons
at tree level approximation in \cite{Georgiou:2010mf}, as well as their one-loop contribution into the Callan-Symanzik beta function \cite{Savvidy:2014hha}. This contribution is negative and
corresponds to an asymptotically free theory.  The review article \cite{Savvidy:2015jgv} contains the details of the calculations.
 
I would like to thank the CERN Theory Division, where part 
of this work was completed, for kind hospitality.

\section{\it Appendix }
The explicit expression for the gauge fields transformation is
 \cite{Savvidy:2005fi,Savvidy:2005zm,Savvidy:2005ki}:
\beqa\label{polygauge1}
A^{'}_{\mu} &\rightarrow & A_{\mu} +  \nabla_{\mu}\xi ,~~~~~\\
 A^{'}_{\mu\lambda_1} &\rightarrow& A_{\mu\lambda_1} + \nabla_{\mu} \xi_{\lambda_1}
-i   g [A_{\mu},  \xi_{\lambda_1}] -  i g [A_{\mu\lambda_1},\xi] ,\nonumber\\
 A^{'}_{\mu\lambda_1 \lambda_2}& \rightarrow&  A_{\mu\lambda_1 \lambda_2} +  \nabla_{\mu} \xi_{\lambda_1\lambda_2} 
-i g (  [A_{\mu \lambda_1}, \xi_{\lambda_2 }] +
[A_{\mu \lambda_2 }, \xi_{\lambda_1}]+
[A_{\mu\lambda_1\lambda_2}, \xi]),\nn\\
.........&.&............................ \nn
\eeqa
and the  components  of the  field strength  tensor are
\cite{Savvidy:2005fi,Savvidy:2005zm,Savvidy:2005ki}:
\beqa\label{fieldstrengthparticular}
G^{a}_{\mu\nu} &=&
\partial_{\mu} A^{a}_{\nu} - \partial_{\nu} A^{a}_{\mu} +
g f^{abc}~A^{b}_{\mu}~A^{c}_{\nu},\\
G^{a}_{\mu\nu,\lambda} &=&
\partial_{\mu} A^{a}_{\nu\lambda} - \partial_{\nu} A^{a}_{\mu\lambda} +
g f^{abc}(~A^{b}_{\mu}~A^{c}_{\nu\lambda} + A^{b}_{\mu\lambda}~A^{c}_{\nu} ~),\nn\\
G^{a}_{\mu\nu,\lambda\rho} &=&
\partial_{\mu} A^{a}_{\nu\lambda\rho} - \partial_{\nu} A^{a}_{\mu\lambda\rho} +
g f^{abc}(~A^{b}_{\mu}~A^{c}_{\nu\lambda\rho} +
 A^{b}_{\mu\lambda}~A^{c}_{\nu\rho}+A^{b}_{\mu\rho}~A^{c}_{\nu\lambda}
 + A^{b}_{\mu\lambda\rho}~A^{c}_{\nu} ~),\nn\\
 ......&.&............................................\nn
\eeqa
The transversal tensor  gauge fields $A_{\mu  +\cdot\cdot\cdot-}$ appear in the expansion  (\ref{gaugefield2}): 
 \beqa
\CA(x,e)= A_{\mu} &+ A_{\mu \lambda_1} \chi n^{\lambda_1} +&{1\over 2}
A_{\mu \lambda_1\lambda_2} (\chi^2 n^{\lambda_1} n^{\lambda_2} + 
e^{\lambda_1}_{+} e^{\lambda_2}_{-} +e^{\lambda_1}_{-} e^{\lambda_2}_{+} )+...\nn\\
&+A_{\mu \lambda_1} e^{i \varphi} e^{\lambda_1}_{+} +&{1\over 2} A_{\mu \lambda_1\lambda_2}  e^{i \varphi} ( 
 e^{\lambda_1}_{+} \chi n^{\lambda_2}  + \chi n^{\lambda_1}   e^{\lambda_2}_{+} )+...\nn\\
&+A_{\mu \lambda_1} e^{-i \varphi} e^{\lambda_1}_{-} +&{1\over 2} A_{\mu \lambda_1\lambda_2}  e^{-i \varphi} ( 
 e^{\lambda_1}_{-} \chi n^{\lambda_2}  + \chi n^{\lambda_1}   e^{\lambda_2}_{-} )+...\nn\\ 
 &~~~~~~~~~~+&{1\over 2} A_{\mu \lambda_1\lambda_2}  e^{2i \varphi}  
 e^{\lambda_1}_{+ } e^{\lambda_2}_{+ }  +  {1\over 2} A_{\mu \lambda_1\lambda_2} e^{-2i \varphi}  e^{\lambda_1}_{- }e^{\lambda_2}_{-} +...\nn\\
=>   A_{\mu} &+ A_{\mu +}  +A_{\mu -}   +&{1\over 2} A_{\mu ++ } +{1\over 2} A_{\mu --}~~~~  .....
\eeqa
The phase factor $e^{i h \varphi}$ defines the helicity $h$ of the 
generators $L^{\lambda_1 ... \lambda_{s}}_{a +\cdot\cdot\cdot-}= e^{i h \varphi} 
e^{\lambda_1}_{+ }....e^{\lambda_s}_{-} \otimes L_{a} $  .

\end{document}